\documentclass[sigconf]{acmart}
\usepackage{makecell}

\AtBeginDocument{%
  \providecommand\BibTeX{{%
    \normalfont B\kern-0.5em{\scshape i\kern-0.25em b}\kern-0.8em\TeX}}}

\setcopyright{acmcopyright}
\copyrightyear{2022}
\acmYear{2022}

\acmConference[KDD'22]{the ACM
International KDD Conference}{Aug, 2022}{Washington D.C., USA}

\begin{document}

\title{A Semantic Alignment System for Multilingual Query-Product Retrieval}
\subtitle{The first-place entry for Query-Product Ranking of ESCI Challenge at KDD Cup 2022}


\author{Qi Zhang, Zijian Yang, Yilun Huang, Ze Chen, Zijian Cai, Kangxu Wang, Jiewen Zheng, Jiarong He, Jin Gao}
\authornote{Corresponding author}
\email{{zhangqi21,yangzijian,huangyilun,jackchen,caizijian01,wangkangxu,zhengjiewen,gzhejiarong,jgao}@corp.netease.com}
\affiliation{%
  \institution{Interactive Entertainment Group of Netease Inc.}
  \city{Guangzhou}
  \country{China}
}

% \author{Qi Zhang}
% \affiliation{%
%   \institution{Interactive Entertainment Group of Netease Inc.}
%   \city{Guangzhou}
%   \country{China}
% }
% \email{zhangqi21@corp.netease.com}

% \author{Zijian Yang}
% \affiliation{%
%   \institution{Interactive Entertainment Group of Netease Inc.}
%   \city{Guangzhou}
%   \country{China}
% }
% \email{yangzijian@corp.netease.com}

% \author{Yilun Huang}
% \affiliation{%
%   \institution{Interactive Entertainment Group of Netease Inc.}
%   \city{Guangzhou}
%   \country{China}
% }
% \email{huangyilun@corp.netease.com}

% \author{Jiarong He}
% \authornote{Corresponding author}
% \affiliation{%
%   \institution{Interactive Entertainment Group of Netease Inc.}
%   \city{Guangzhou}
%   \country{China}
% }
% \email{gzhejiarong@corp.netease.com}


\begin{abstract}

This paper mainly describes our winning solution (team name: \textit{www}) to Amazon ESCI Challenge of KDD CUP 2022, which achieves a NDCG score of 0.9043 and wins the first place on task 1: the query-product ranking track.\footnote{Final Winners Announcement: https://discourse.aicrowd.com/t/final-winners-announcement}

In this competition, participants are provided with a real-world large-scale multilingual shopping queries data set and it contains query-product pairs in English, Japanese and Spanish. Three different tasks are proposed in this competition, including ranking the results list as task 1, classifying the query/product pairs into Exact, Substitute, Complement, or Irrelevant (ESCI) categories as task 2 and identifying substitute products for a given query as task 3. 

We mainly focus on task 1 and propose a semantic alignment system for multilingual query-product retrieval. Pre-trained multilingual language models (LM) are adopted to get the semantic representation of  queries and products. Our models are all trained with cross-entropy loss to classify the query-product pairs into ESCI  4 categories at first, and then we use weighted sum with the 4-class probabilities to get the score for ranking. To further boost the model, we also do elaborative data preprocessing, data augmentation by translation, specially handling English texts with English LMs, adversarial training with AWP and FGM, self distillation, pseudo labeling, label smoothing and ensemble. Finally, Our solution outperforms others both on public and private leaderboard.

\end{abstract}


% %% A "teaser" image appears between the author and affiliation
% %% information and the body of the document, and typically spans the

\maketitle

\section{Introduction}
Amazon ESCI Challenge  \cite{U1} for Improving Product Search of KDD CUP 2022 is aiming to improve the customer experience and their engagement when searching for products. The primary objective of this competition is to build new ranking strategies and, simultaneously, to identify interesting categories of results by using their real-world Shopping Queries Dataset. 

\subsection{Dataset Description}

The provided Shopping Queries Dataset \cite{reddy2022shopping} involves 3 languages: English (about 54.5\% of the total training sets) , Japanese (about 26.5\% of the total training sets) and Spanish (about 19\% of the total training sets). In online shopping applications, the notion of binary relevance limits the customer experience. To keep high accuracy in ranking, the competition organizers break down relevance into the following four classes (ESCI) which are used to measure the relevance of the items in the search results.  Exact (E) and Substitute (S), stands for the item is relevant and somewhat relevant to the query respectively.  Complement (C) and Irrelevant (I) denotes respectively the item does not fulfill the query and the item is irrelevant. 

For each query, the dataset provides a list of up to 40 potentially relevant product results, together with ESCI relevance judgements and an annotated locale label. For each product,  the dataset provides associated information such as product title, description, bullet points, brand, color and locale.

A total of about 1.2 million products and 2 million query-product pairs are provided in this challenge, which are used for model training and offline validating. Online test set is split into public and private ones, and the final ranking is based on the score on the private leaderboard.  

As shown in Table 1, the label distribution is very imbalanced. Most of the labels are Exact with the percentage up to 62.78\%, while Complement class only accounts for 3.16\%, Substitute and Irrelevant class account for 23.28\% and 10.78\% respectively. And according to our statistics, 54\% of product brands focus on providing only one product, while only 7.1\% of brands provide more than 10 products. At the same time, more than 80\% of the color names are only customized for a single product. Such results help us to further understand and quantify the characteristics and distributions of the corpora provided in this competition.

\subsection{Task Description}

There are three tasks in this competition and we mainly involved in task 1: Query-Product Ranking. The goal of this task is to rank a list of matched products of a specified user query. NDCG is used as relevance metric in this task, with a class gain of 1.0, 0.1, 0.01, 0.0 setting for ESCI respectively. The input for this task is a list of queries with their product identifiers.  And the participants is asked to build a system to sort the product candidates, with the most relevant product in the first row and the least relevant product in the last.

\begin{table}
  \caption{Distribution of ESCI Labels}
%   \label{tab:commands}
  \begin{tabular}{lc}
    \toprule
    \# (E) Exact & 62.78\% \\
    \# (S) Substitute & 23.28\% \\
    \# (C) Complement & 3.16\% \\
    \# (I) Irrelevant & 10.78\% \\
    \bottomrule
  \end{tabular}
\end{table}

\section{Related Work}

Our work is mainly related to the pre-trained LMs and some specific strategies on Learning-to-Rank(LTR) systems, such as adversarial training, model ensemble and so on.

\subsection{Cross-Encoder Models}
Neural approaches have greatly improved the information retrieval results in recent years. Prior to this, similarity metrics primarily rely on keyword matching, with some limited thesaurus and phrase-based expansion. BERT\cite{devlin2018bert,abolghasemi2022improving} uses Cross-Encoder architecture to achieve further improvements in the field of text understanding by passing the query and product simultaneously to the transformer networks and producing an output representation that indicates the similarity of the input pairs. 

\subsection{Multilingual Language Model}
Defining textual features in a cross-lingual representation space has always been a challenge. The more languages there are, the confusing the representation contents will be. XLM \cite{lample2019cross} use Byte-Pair Encoding that splits the input into the most common sub-words across different languages instead of using word or characters as the input of the model. On the other hand, the Translation Language Modeling (TLM) task also increases the ability of contextual encoding. Nowadays, a set of large-scale Transformer-based pre-trained language models, such as RemBERT\cite{chung2020rethinking}, XLM-RoBERTa\cite{conneau2019unsupervised}, InfoXLM\cite{chi2020infoxlm} and mDeBERTa\cite{he2021debertav3} have created new state of the art in many downstream fields. It turns out that training cross-lingual language models can improve performance on many NLP tasks.

\subsection{Learning to Rank}
Learning to rank (LTR) is a class of algorithmic techniques which are used to solve ranking problems in search relevancy\cite{cao2007learning}. For a specific query, we can use the model to do the ranking so that the relevant products will be ranked above the non-relevant ones. During the modeling process, query embedding and product embedding are concatenated as input to refine the self-attention mechanism for learning semantic alignments between queries and product descriptions. It is significant for the ranking task to establish a strong semantic alignment between the queries and the products\cite{liu2021que2search}.

\section{Methodology}

Our solution to this task mainly consists of 3 parts, which are data preprocessing, model training and ensemble. The overall framework is shown in Figure 1.

\begin{figure}
    \includegraphics[width=\linewidth]{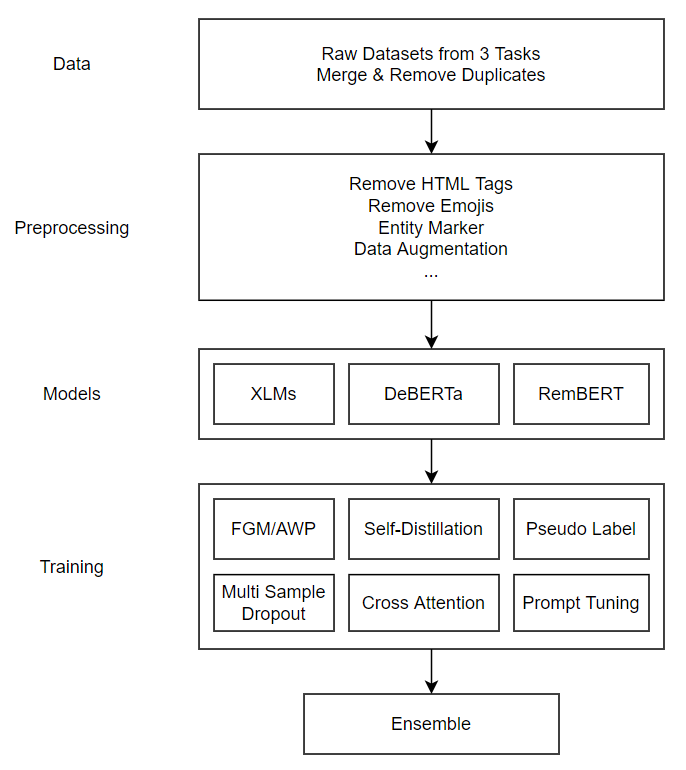}
    \caption{An overall framework and pipeline of our solution}
    \label{fig:fig4}
\end{figure}

\subsection{Data Processing}

Given that datasets from the 3 tasks are exactly in the same format, we use all of the data from 3 tasks to train the model for task 1. The raw data is quite noisy containing many useless html tags, symbols and emojis, so we do some data cleaning work before model training by simply remove these trivial stuffs. After data cleaning, we use Google API to translate all of the data into English, Spanish and Japanese separately to do data augmentation.

We also use typed entity marker\cite{zhong2020frustratingly} to incorporate the NER information into the input of models. We add special tokens [TYPE], [/TYPE] near the entities in input text, where TYPE is the entity type recognized by a named entity tagger. For example, given the query “I want to buy an iPhone 8 Plus”, it will be modified to “I want to buy an [Product] iPhone 8 Plus [/Product]”.

\subsection{Model Architecture}

The single model architecture is shown in the Figure 2, we use the cross-encoder architecture based on DeBERTa, XLMs, and RemBERT. For the downstream task, [CLS] embedding is used to do the ESCI 4-class classification. We concatenate the query and product information in the way of “[CLS]query[SEP]color:<color> brand:<brand> description:<title+bullet\_point+description>[SEP]” before feeding into the models. 

\begin{figure}
    \includegraphics[width=\linewidth]{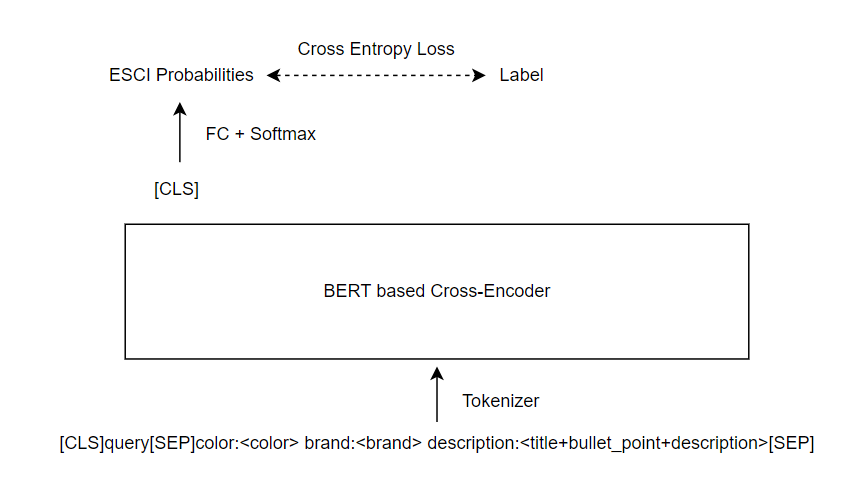}
    \caption{Model architecture for a single model}
    \label{fig:fig4}
\end{figure}

\subsection{Training Details}

We found that lots of mislabeling cases may exist in the dataset when doing bad cases analysis based on our local validation set in ESCI classification task. We realize that the data, labeled by crowdsourcing, could be quite noisy inevitably. To avoid misleading by the mislabeling, we use label smoothing to make the model less confident to the labels, which is proven to be effective. Training models in complex and ambiguous contexts can badly affects its generalization. To improve the model robustness, some training tricks are adopted during our experiments. 

\subsubsection{Self Distillation\cite{xu2020improving}}

Given that the data is labeled by crowdsourcing, it could be quite noisy for the model to extract the real information. To make the model more robust, we use self-distillation training to further boost our single model. The model itself is used as its own teacher to do distillation training. To be specific, we use 3-fold bagging training and make prediction on the out-of-fold datasets to generate the soft labels for all of the training examples. And then we merge the soft labels with the ground true hard labels with weights 0.3 and 0.7 to get the new training labels: $y\_new = 0.7 * hard\ \ labels + 0.3*soft\ \ labels$

We also tried to use two loss functions to compute the soft label loss and hard label loss separately and sum it with different weights. Unfortunately, it didn't work better than directly merge the labels as mentioned above.

\subsubsection{Pseudo Label\cite{lee2013pseudo}}
We also use our trained models to generate pseudo labels from the public test set to do further training. To avoid making the training data more noisy, only samples from the public test set with predicted probabilities above 0.7 are used as pseudo labels, as shown in Figure 3. And soft labels work better than hard labels during most of our experiments, we guess that hard labels may increase the risk of overfitting to some extent.

\begin{figure}
    \includegraphics[width=\linewidth]{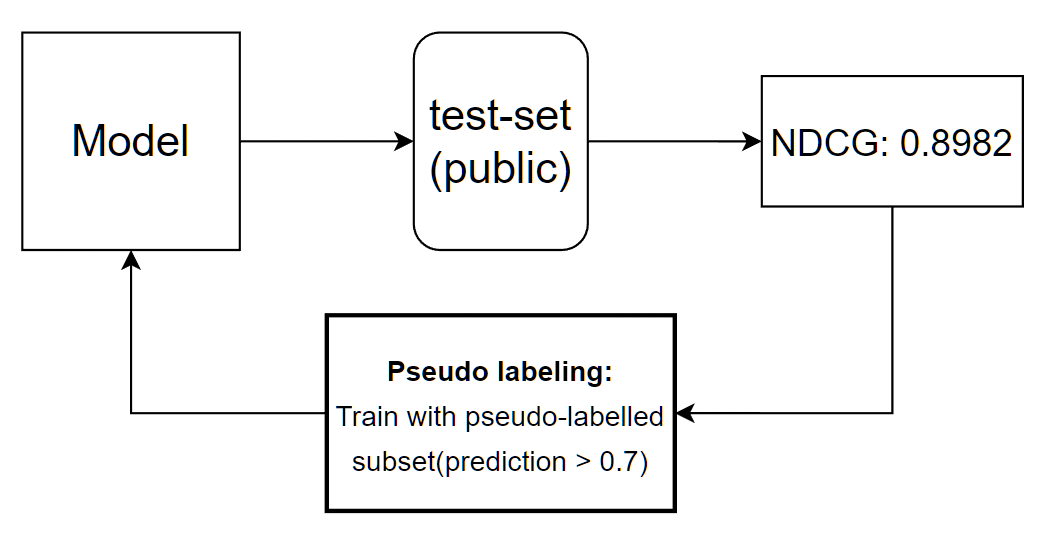}
    \caption{Train model with pseudo-labelled subset}
    \label{fig:fig1}
\end{figure}

\subsubsection{Multi Sample Dropout\cite{inoue2019multi}}

Dropout is a simple but efficient regularization technique for achieving better generalization. By combining a group of dropout layers with different dropout ratios, multi-sample dropout can achieve further improvement for the model. In this scenario, we use multi-sample dropout before the output layer to make our model more robust. 

\subsubsection{Prompt Tuning\cite{lester2021power}}
In addition to increasing the robustness and recall of the model, we also make some adjustments to improve the precision. On the step of data processing, we prepared a series of specific tokens as implicit templates to provide extended reference features during encoding. By introducing this information, our model can have a better performance to handle diverse features. 

\subsubsection{Cross Attention\cite{cui2016attention}}

Furthermore, it is obviously to found that the textual information between the query and the content of product are different in most cases (the text length of the product is much longer than the query). In order to prevent the query information vanishing after the neural transmission, our model automatically generates an attended attention over original self-attention, which makes the feature of latent distribution not only contain the query-to-product part, but also the query-to-mix\_sequence. 

\subsubsection{Adversarial Training}
Adversarial-training let us train networks with significantly improved resistance to adversarial attacks, thus improving robustness of models. 

When the loss is below some threshold (like 0.6), we start using Adversarial Weight Perturbation (AWP) \cite{wu2020adversarial} in training steps that adversarially perturbs both model weights and the embeddings. In addition, the feature distribution of input data is attacked in each step. Besides, we also tried Fast Gradient Method (FGM) \cite{goodfellow2014explaining} which performs slightly worse than AWP does in public leaderboard.

\subsubsection{English BERT Model}
Although the task is multilingual, the English part is of large proportion, accounting for 54.5\% of the total training sets. Take this into account, we also use DeBERTa-v3-large to train and predict for the English queries and products only besides the cross-lingual models. Combining with adversarial training, our single model gets improved from 0.899 to 0.9022 in the public leaderboard.

\subsection{Ensemble}
At last, model ensemble is used to get the final improvement. In detail, we use DeBERTa, RemBERT and XLM based models trained with different settings mentioned above as our base models for ensemble. 

The weights for summing different model predictions are mainly determined by the public scores of the models and also the local cross-validation scores. We also lower the weights of the models with high correlation coefficients. Our score is improved from 0.9022 to 0.9057 on the public leaderboard, and from 0.9015 to 0.9043 on the private leaderboard after ensemble.

\section{Results \& Discussion}

Some results of our experiments in Task 1 are shown in Table 2. The scores of our single models without any pre-processing or post-processing are around 0.8930 in the public leaderboard. DeBERTa, InfoXLM, XLM-RoBERTa, and RemBERT are used as the model backbone, then we concatenate the texts of query, title, description, bullet point together and truncate it with \verb|max_length=128| after tokenizing as the model inputs. 

After doing some data cleaning and hyper-parameters tuning, our single model is improved to 0.8960 on the public leaderboard. Batch size and learning rate are quite important in this task based on our experiments, we use \verb|batch size=64|, \verb|learning rate=3e-5| and \verb|gradient accumulation=8| to train the model after tuning. Self distillation, pseudo labels and label smoothing help us further boost the model performance to 0.8990 on the public leaderboard.

With English pre-trained LMs and adversarial training, we can achieve 0.9022 in the public leaderboard and 0.9015 for the private, and this is our best single model. At last, we do model ensemble to get the final boost from 0.9022 to 0.9057 on the public leaderboard, and from 0.9015 to 0.9043 on the private leaderboard.

\begin{table}[!ht]
    \caption{Some results of our experiments in Task 1}
    \centering
    \begin{tabular}{p{4.5cm} p{1.3cm} p{1.3cm}}
    \hline
        Methods & NDCG (Public) & NDCG (Private) \\ 
        \hline
        mDeBERTa Baseline & 0.8930 & - \\ 
        \hline
        \makecell[l]{+ Data Clean \\ + Parameter Tuning} & \makecell[l]{0.8960} & \makecell[l]{-} \\ 
        \hline
        \makecell[l]{+ Self Distillation  \\ + Pseudo Labeling} & 0.8982 & 0.8975 \\ 
        \hline
        + Label Smoothing & 0.8990 & - \\ 
        \hline
        \makecell[l]{+ DeBERTa-v3-large \\ + AWP/FGM} & 0.9022 & 0.9015  \\ 
        \hline
        + InfoXLM & 0.9032 & 0.9032  \\ 
        \hline
        \makecell[l]{+ XLM-RoBERTa } & 0.9041 & 0.9026 \\ 
        \hline
        \makecell[l]{+ RemBERT \\ + DeBERTa-v3-large \\ \ \ + Translation augmentation \\
        \ \ + 2 of 7 folds bagging \\ \ \ + Weighted multi-layer Pooling \\ 
        \ \ + Multi sample dropout}  &  \textbf{0.9059*} &  0.9039 \\ 
        \hline
        Model Ensemble Re-weighting & 0.9057 & \textbf{0.9043*}  \\ 
        \hline
    \end{tabular}
\end{table}

\section{Conclusion}

In this paper, we detailed our winning solution to the Query-Product Ranking task in Amazon ESCI Challenge of KDD Cup 2022. We use multilingual and English pre-trained LMs as backbone, with the combination of data processing, data augmentation, self-distillation, pseudo-labelling, label-smoothing and adversarial training, we improve the model step by step. For single model, we achieve NDCG score of 0.9022 on the public leaderboard and 0.9015 on the private leaderboard. At last, we do model ensemble to get  the final boost from 0.9022 to 0.9057 on the public leaderboard, and from 0.9015 to 0.9043 on the private leaderboard, which ensures us to 
win the first place. 

\begin{acks}
Amazon and AIcrowd organizing team paid a lot of efforts during the whole process of the competition, we really appreciate it for hosting this fantastic competition. And we would like to thank everyone associated with organizing and sponsoring the KDD Cup 2022.
\end{acks}

\bibliographystyle{ACM-Reference-Format}
\bibliography{sample-base}


\begin{thebibliography}{19}

%%% the Web address.

\ifx \showCODEN    \undefined \def \showCODEN     #1{\unskip}     \fi
\ifx \showDOI      \undefined \def \showDOI       #1{#1}\fi
\ifx \showISBNx    \undefined \def \showISBNx     #1{\unskip}     \fi
\ifx \showISBNxiii \undefined \def \showISBNxiii  #1{\unskip}     \fi
\ifx \showISSN     \undefined \def \showISSN      #1{\unskip}     \fi
\ifx \showLCCN     \undefined \def \showLCCN      #1{\unskip}     \fi
\ifx \shownote     \undefined \def \shownote      #1{#1}          \fi
\ifx \showarticletitle \undefined \def \showarticletitle #1{#1}   \fi
\ifx \showURL      \undefined \def \showURL       {\relax}        \fi
% invisible to TeX
\providecommand\bibfield[2]{#2}
\providecommand\bibinfo[2]{#2}
\providecommand\natexlab[1]{#1}
\providecommand\showeprint[2][]{arXiv:#2}

\bibitem[Abolghasemi et~al\mbox{.}(2022)]%
        {abolghasemi2022improving}
\bibfield{author}{\bibinfo{person}{Amin Abolghasemi}, \bibinfo{person}{Suzan
  Verberne}, {and} \bibinfo{person}{Leif Azzopardi}.}
  \bibinfo{year}{2022}\natexlab{}.
\newblock \showarticletitle{Improving BERT-based query-by-document retrieval
  with multi-task optimization}. In \bibinfo{booktitle}{\emph{European
  Conference on Information Retrieval}}. Springer, \bibinfo{pages}{3--12}.
\newblock


\bibitem[{aicrowd}(2022)]%
        {U1}
\bibfield{author}{\bibinfo{person}{{aicrowd}}.}
  \bibinfo{year}{2022}\natexlab{}.
\newblock \bibinfo{title}{ESCI Challenge for Improving Product Search}.
\newblock
  \bibinfo{howpublished}{\url{https://www.aicrowd.com/challenges/esci-challenge-for-improving-product-search}}.
\newblock


\bibitem[Cao et~al\mbox{.}(2007)]%
        {cao2007learning}
\bibfield{author}{\bibinfo{person}{Zhe Cao}, \bibinfo{person}{Tao Qin},
  \bibinfo{person}{Tie-Yan Liu}, \bibinfo{person}{Ming-Feng Tsai}, {and}
  \bibinfo{person}{Hang Li}.} \bibinfo{year}{2007}\natexlab{}.
\newblock \showarticletitle{Learning to rank: from pairwise approach to
  listwise approach}. In \bibinfo{booktitle}{\emph{Proceedings of the 24th
  international conference on Machine learning}}. \bibinfo{pages}{129--136}.
\newblock


\bibitem[Chi et~al\mbox{.}(2020)]%
        {chi2020infoxlm}
\bibfield{author}{\bibinfo{person}{Zewen Chi}, \bibinfo{person}{Li Dong},
  \bibinfo{person}{Furu Wei}, \bibinfo{person}{Nan Yang},
  \bibinfo{person}{Saksham Singhal}, \bibinfo{person}{Wenhui Wang},
  \bibinfo{person}{Xia Song}, \bibinfo{person}{Xian-Ling Mao},
  \bibinfo{person}{Heyan Huang}, {and} \bibinfo{person}{Ming Zhou}.}
  \bibinfo{year}{2020}\natexlab{}.
\newblock \showarticletitle{InfoXLM: An information-theoretic framework for
  cross-lingual language model pre-training}.
\newblock \bibinfo{journal}{\emph{arXiv preprint arXiv:2007.07834}}
  (\bibinfo{year}{2020}).
\newblock


\bibitem[Chung et~al\mbox{.}(2020)]%
        {chung2020rethinking}
\bibfield{author}{\bibinfo{person}{Hyung~Won Chung}, \bibinfo{person}{Thibault
  Fevry}, \bibinfo{person}{Henry Tsai}, \bibinfo{person}{Melvin Johnson}, {and}
  \bibinfo{person}{Sebastian Ruder}.} \bibinfo{year}{2020}\natexlab{}.
\newblock \showarticletitle{Rethinking embedding coupling in pre-trained
  language models}.
\newblock \bibinfo{journal}{\emph{arXiv preprint arXiv:2010.12821}}
  (\bibinfo{year}{2020}).
\newblock


\bibitem[Conneau et~al\mbox{.}(2019)]%
        {conneau2019unsupervised}
\bibfield{author}{\bibinfo{person}{Alexis Conneau}, \bibinfo{person}{Kartikay
  Khandelwal}, \bibinfo{person}{Naman Goyal}, \bibinfo{person}{Vishrav
  Chaudhary}, \bibinfo{person}{Guillaume Wenzek}, \bibinfo{person}{Francisco
  Guzm{\'a}n}, \bibinfo{person}{Edouard Grave}, \bibinfo{person}{Myle Ott},
  \bibinfo{person}{Luke Zettlemoyer}, {and} \bibinfo{person}{Veselin
  Stoyanov}.} \bibinfo{year}{2019}\natexlab{}.
\newblock \showarticletitle{Unsupervised cross-lingual representation learning
  at scale}.
\newblock \bibinfo{journal}{\emph{arXiv preprint arXiv:1911.02116}}
  (\bibinfo{year}{2019}).
\newblock


\bibitem[Cui et~al\mbox{.}(2016)]%
        {cui2016attention}
\bibfield{author}{\bibinfo{person}{Yiming Cui}, \bibinfo{person}{Zhipeng Chen},
  \bibinfo{person}{Si Wei}, \bibinfo{person}{Shijin Wang},
  \bibinfo{person}{Ting Liu}, {and} \bibinfo{person}{Guoping Hu}.}
  \bibinfo{year}{2016}\natexlab{}.
\newblock \showarticletitle{Attention-over-attention neural networks for
  reading comprehension}.
\newblock \bibinfo{journal}{\emph{arXiv preprint arXiv:1607.04423}}
  (\bibinfo{year}{2016}).
\newblock


\bibitem[Devlin et~al\mbox{.}(2018)]%
        {devlin2018bert}
\bibfield{author}{\bibinfo{person}{Jacob Devlin}, \bibinfo{person}{Ming-Wei
  Chang}, \bibinfo{person}{Kenton Lee}, {and} \bibinfo{person}{Kristina
  Toutanova}.} \bibinfo{year}{2018}\natexlab{}.
\newblock \showarticletitle{Bert: Pre-training of deep bidirectional
  transformers for language understanding}.
\newblock \bibinfo{journal}{\emph{arXiv preprint arXiv:1810.04805}}
  (\bibinfo{year}{2018}).
\newblock


\bibitem[Goodfellow et~al\mbox{.}(2014)]%
        {goodfellow2014explaining}
\bibfield{author}{\bibinfo{person}{Ian~J Goodfellow}, \bibinfo{person}{Jonathon
  Shlens}, {and} \bibinfo{person}{Christian Szegedy}.}
  \bibinfo{year}{2014}\natexlab{}.
\newblock \showarticletitle{Explaining and harnessing adversarial examples}.
\newblock \bibinfo{journal}{\emph{arXiv preprint arXiv:1412.6572}}
  (\bibinfo{year}{2014}).
\newblock


\bibitem[He et~al\mbox{.}(2021)]%
        {he2021debertav3}
\bibfield{author}{\bibinfo{person}{Pengcheng He}, \bibinfo{person}{Jianfeng
  Gao}, {and} \bibinfo{person}{Weizhu Chen}.} \bibinfo{year}{2021}\natexlab{}.
\newblock \showarticletitle{Debertav3: Improving deberta using electra-style
  pre-training with gradient-disentangled embedding sharing}.
\newblock \bibinfo{journal}{\emph{arXiv preprint arXiv:2111.09543}}
  (\bibinfo{year}{2021}).
\newblock


\bibitem[Inoue(2019)]%
        {inoue2019multi}
\bibfield{author}{\bibinfo{person}{Hiroshi Inoue}.}
  \bibinfo{year}{2019}\natexlab{}.
\newblock \showarticletitle{Multi-sample dropout for accelerated training and
  better generalization}.
\newblock \bibinfo{journal}{\emph{arXiv preprint arXiv:1905.09788}}
  (\bibinfo{year}{2019}).
\newblock


\bibitem[Lample and Conneau(2019)]%
        {lample2019cross}
\bibfield{author}{\bibinfo{person}{Guillaume Lample} {and}
  \bibinfo{person}{Alexis Conneau}.} \bibinfo{year}{2019}\natexlab{}.
\newblock \showarticletitle{Cross-lingual language model pretraining}.
\newblock \bibinfo{journal}{\emph{arXiv preprint arXiv:1901.07291}}
  (\bibinfo{year}{2019}).
\newblock


\bibitem[Lee et~al\mbox{.}(2013)]%
        {lee2013pseudo}
\bibfield{author}{\bibinfo{person}{Dong-Hyun Lee} {et~al\mbox{.}}}
  \bibinfo{year}{2013}\natexlab{}.
\newblock \showarticletitle{Pseudo-label: The simple and efficient
  semi-supervised learning method for deep neural networks}. In
  \bibinfo{booktitle}{\emph{Workshop on challenges in representation learning,
  ICML}}, Vol.~\bibinfo{volume}{3}. \bibinfo{pages}{896}.
\newblock


\bibitem[Lester et~al\mbox{.}(2021)]%
        {lester2021power}
\bibfield{author}{\bibinfo{person}{Brian Lester}, \bibinfo{person}{Rami
  Al-Rfou}, {and} \bibinfo{person}{Noah Constant}.}
  \bibinfo{year}{2021}\natexlab{}.
\newblock \showarticletitle{The power of scale for parameter-efficient prompt
  tuning}.
\newblock \bibinfo{journal}{\emph{arXiv preprint arXiv:2104.08691}}
  (\bibinfo{year}{2021}).
\newblock


\bibitem[Liu et~al\mbox{.}(2021)]%
        {liu2021que2search}
\bibfield{author}{\bibinfo{person}{Yiqun Liu}, \bibinfo{person}{Kaushik
  Rangadurai}, \bibinfo{person}{Yunzhong He}, \bibinfo{person}{Siddarth
  Malreddy}, \bibinfo{person}{Xunlong Gui}, \bibinfo{person}{Xiaoyi Liu}, {and}
  \bibinfo{person}{Fedor Borisyuk}.} \bibinfo{year}{2021}\natexlab{}.
\newblock \showarticletitle{Que2Search: fast and accurate query and document
  understanding for search at Facebook}. In
  \bibinfo{booktitle}{\emph{Proceedings of the 27th ACM SIGKDD Conference on
  Knowledge Discovery \& Data Mining}}. \bibinfo{pages}{3376--3384}.
\newblock


\bibitem[Reddy et~al\mbox{.}(2022)]%
        {reddy2022shopping}
\bibfield{author}{\bibinfo{person}{Chandan~K Reddy},
  \bibinfo{person}{Llu{\'\i}s M{\`a}rquez}, \bibinfo{person}{Fran Valero},
  \bibinfo{person}{Nikhil Rao}, \bibinfo{person}{Hugo Zaragoza},
  \bibinfo{person}{Sambaran Bandyopadhyay}, \bibinfo{person}{Arnab Biswas},
  \bibinfo{person}{Anlu Xing}, {and} \bibinfo{person}{Karthik Subbian}.}
  \bibinfo{year}{2022}\natexlab{}.
\newblock \showarticletitle{Shopping Queries Dataset: A Large-Scale ESCI
  Benchmark for Improving Product Search}.
\newblock \bibinfo{journal}{\emph{arXiv preprint arXiv:2206.06588}}
  (\bibinfo{year}{2022}).
\newblock


\bibitem[Wu et~al\mbox{.}(2020)]%
        {wu2020adversarial}
\bibfield{author}{\bibinfo{person}{Dongxian Wu}, \bibinfo{person}{Shu-Tao Xia},
  {and} \bibinfo{person}{Yisen Wang}.} \bibinfo{year}{2020}\natexlab{}.
\newblock \showarticletitle{Adversarial weight perturbation helps robust
  generalization}.
\newblock \bibinfo{journal}{\emph{Advances in Neural Information Processing
  Systems}}  \bibinfo{volume}{33} (\bibinfo{year}{2020}),
  \bibinfo{pages}{2958--2969}.
\newblock


\bibitem[Xu et~al\mbox{.}(2020)]%
        {xu2020improving}
\bibfield{author}{\bibinfo{person}{Yige Xu}, \bibinfo{person}{Xipeng Qiu},
  \bibinfo{person}{Ligao Zhou}, {and} \bibinfo{person}{Xuanjing Huang}.}
  \bibinfo{year}{2020}\natexlab{}.
\newblock \showarticletitle{Improving bert fine-tuning via self-ensemble and
  self-distillation}.
\newblock \bibinfo{journal}{\emph{arXiv preprint arXiv:2002.10345}}
  (\bibinfo{year}{2020}).
\newblock


\bibitem[Zhong and Chen(2020)]%
        {zhong2020frustratingly}
\bibfield{author}{\bibinfo{person}{Zexuan Zhong} {and} \bibinfo{person}{Danqi
  Chen}.} \bibinfo{year}{2020}\natexlab{}.
\newblock \showarticletitle{A frustratingly easy approach for entity and
  relation extraction}.
\newblock \bibinfo{journal}{\emph{arXiv preprint arXiv:2010.12812}}
  (\bibinfo{year}{2020}).
\newblock


\end{thebibliography}

% \appendix









\end{document}